\documentclass[fleqn,10pt]{wlscirep}
\usepackage[utf8]{inputenc}
\usepackage[T1]{fontenc}
\usepackage{xurl}

\newcommand{\Ivan}{\color{black}}

\title{Exploring Cognitive Paradoxes in Video Games: A Quantum Mechanical Perspective}

\author[1,*]{Ivan S.~Maksymov}
\author[1,2,3]{Ganna~Pogrebna}
\affil[1]{Artificial Intelligence and Cyber Futures Institute, Charles Sturt University, Bathurst, NSW 2795, Australia}
\affil[2]{The Alan Turing Institute, British Library, 96 Euston Rd., London NW1 2DB, United Kingdom}
\affil[3]{The University of Sydney Business School, Abercrombie Building H70, Corner Abercrombie Street and Codrington St., Darlington NSW 2006, Australia}

\affil[*]{imaksymov@csu.edu.au}

\begin{abstract}
This paper introduces a quantum-mechanical model that bridges the realms of cognition and quantum mechanics, offering a novel perspective on decision-making under risk and perceptual reversals. By integrating quantum theories addressing decision-theoretic anomalies with examples from immersive video games like ``Deal or No Deal'', we seek to elucidate complex human cognitive behaviours. Study~1 showcases the proposed quantum model's superiority over traditional decision-making approaches using the ``Deal or No Deal'' video game experiment. In Study~2, we apply our model to bistable perceptions, taking the Necker cube from the Necker game as a primary example. While previous works have hinted at connections between quantum mechanics and cognition, Study~3 provides a more tangible link, likening the physics that underpins quantum tunnelling to an eye blink's role in perceptual reversals. Conclusively, our model displays a promising ability to interpret diverse optical illusions and psychological phenomena, marking a significant stride in understanding human decision making.

\end{abstract}
\begin{document}

\flushbottom
\maketitle
\thispagestyle{empty}

%\noindent Please note: Abbreviations should be introduced at the first mention in the main text – no abbreviations lists. Suggested structure of main text (not enforced) is provided below.

\section*{Introduction}

The landscape of decision-making research has been marked by intriguing inconsistencies in human choices under risk, often revealing alterations in decisions when subjected to repeated trials \cite{hey1994investigating, blavatskyy2010endowment}. Conventional deterministic decision theories often struggle to account for such alterations \cite{blavatskyy2010models}, which is especially apparent in data obtained from video games \cite{bailey2013would}. This paper proposes a pioneering quantum-mechanical model, weaving connections between psychological phenomena and fundamental principles from the realm of physics, particularly focusing on bistable perception. We also show that this model is capable of explaining observed behavioural regularities better than existing decision theories. Video games provide an intriguing platform for exploring decision-making, where human choices under risk reveal surprising inconsistencies and cognitive paradoxes. Our study leverages this medium to connect quantum mechanics and bistable perception.

Recent times have witnessed a surge in the popularity of video games, offering a promising avenue for the investigation of decision-making processes within a controlled yet lifelike setting. Among those, the video game ``Deal or No Deal'' (released in 2006 in the US and in 2008 in the UK by the Gravity-i after the global success of the Endemol~TV show with the same name) stands out as an exemplary terrain for experimental research on decision-making and risk-taking behaviours. The UK version of the video game repeats the ``Deal or No Deal UK'' show broadcast on Channel~4. 

{\Ivan Each game features 22~contestants represented by avatars holding boxes with monetary prizes. The player can select any avatar and any box number and then play the game opening the remaining boxes one by one. Prizes in the game range from \pounds0.01 to \pounds250,000 (hypothetical money). These amounts are randomly distributed across 22~boxes. Once a box is opened and a prize is revealed, it is eliminated from the list of possible prizes. The goal of the participant is to play the game until the end and win the maximum prize of \pounds250,000. Yet, players rarely play the game until the last box and get monetary offers of sure amounts of money from the ``banker'' after opening specific sequences of boxes: first 5, then 3 repeatedly. Since game players are constantly making choices between a risky lottery and an amount of money for sure (bank offer), the game represents an excellent platform to study human behaviour under risk \cite{pogrebna2008naive,blavatskyy2010models}.

The distinction between sticking and swapping decisions can be evaluated through the lenses of the field of decision theory, particularly, via the application of Expected Utility Theory (EUT see, e.g., Ref.~\cite{Von44}) and Cumulative Prospect Theory (CPT; see, e.g., Ref.~\cite{Tve92}). Under the deterministic version of EUT, if the expected value of swapping is equal to the expected value of sticking, the model predicts indifference between the two choices.

The term ``deterministic decision theories'' is standard in decision theory literature and refers to the theories that predict stable choices under identical conditions. The original formulations of EUT and CPT, as well as of the other foundational models, are deterministic in nature, meaning that a decision-maker who prefers option $A$ over $B$ will always make the same choice when faced with the same decision. However, empirical findings in experimental economics and psychology reveal that individuals frequently exhibit choice variability when confronted with repeated decisions, even in identical settings. This discrepancy has led to the development of stochastic models of choice, which introduce probabilistic elements to account for observed behaviour (see, e.g., Refs.~\cite{fechner1948elements, luce1959individual, blavatskyy2010models}).

For instance, while Fechner's model incorporates random errors \cite{fechner1948elements}, Luce's model introduces probabilistic decision rules \cite{luce1959individual} but rank-dependent stochastic models account for response variability. The distinction between deterministic and stochastic choice models is critical since it highlights the need for additional frameworks that can accommodate empirical inconsistencies in human decision-making \cite{hey1994investigating, loomes2017preference}. However, the introduction of stochastic elements, such as Fechner's and Luce's error models, accounts for variability in participant responses across repeated trials. In contrast, CPT posits that loss aversion plays a dominant role in decision-making, leading individuals to overvalue their current choice and disproportionately fear losses associated with swapping. As a result, CPT predicts a stronger preference for sticking to the original box, even when swapping has an objectively equal or higher expected value. Empirical findings from one of our experiments (described below) indicate that while 63\% of participants chose to stick, the observed choice distribution is continuous rather than binary. This suggests that pure decision-theoretic deterministic models do not fully capture decision behaviour, requiring the application of more sophisticated modelling via embedding these theories into models of stochastic choice. This highlights the necessity of integrating stochastic choice mechanisms into theoretical decision-making frameworks \cite{loomes2017preference}.

Historically, studies on decision-making under risk have illuminated the propensity for individuals to modify their choices when confronted with identical binary decision problems repeated over short spans \cite{loomes2017preference}. Traditional deterministic decision theories, rooted in the assumption of consistent choices upon repetition (such as already described deterministic versions of EUT and even Rank-Dependent Utility, of which CPT is a special case), often fall short of explaining these observed deviations \cite{blavatskyy2010models}. Concurrently, a quantum-mechanical perspective has arisen, introducing new possibilities for understanding psychological phenomena through a blend of physics and human behaviour \cite{Bus12, Pot22}. While previous findings highlighted the significance of decision theory and stochastic choice model selection in influencing the estimation of decision theories \cite{blavatskyy2010models}, we demonstrate that synergies between physics and psychology produce more powerful opportunities to explain observed behaviour.}

Recent decades have seen numerous attempts to address the question whether the unparalleled computing power of a biological brain could be explained by quantum-mechanical processes that are likely to occur in the neurons \cite{Khr06, Koch06, Aer22}. Studies that incorporate physical concepts have also been conducted in the field of psychology \cite{mason2016neural, ludwin2020broken, wong2023seeing}, where the hypotheses of quantum mind and quantum consciousness were proposed \cite{mindell2012quantum, wendt2015quantum} to improve the existing classical models of brain function \cite{Koch06}. As a result of these endeavours, quantum-physical models of cognition and decision-making have been developed \cite{Khr06, Bus12, Khr14, Yea16, Oza20, Pot22, Wad23}.
 \begin{figure}
 \includegraphics[width=0.89\textwidth]{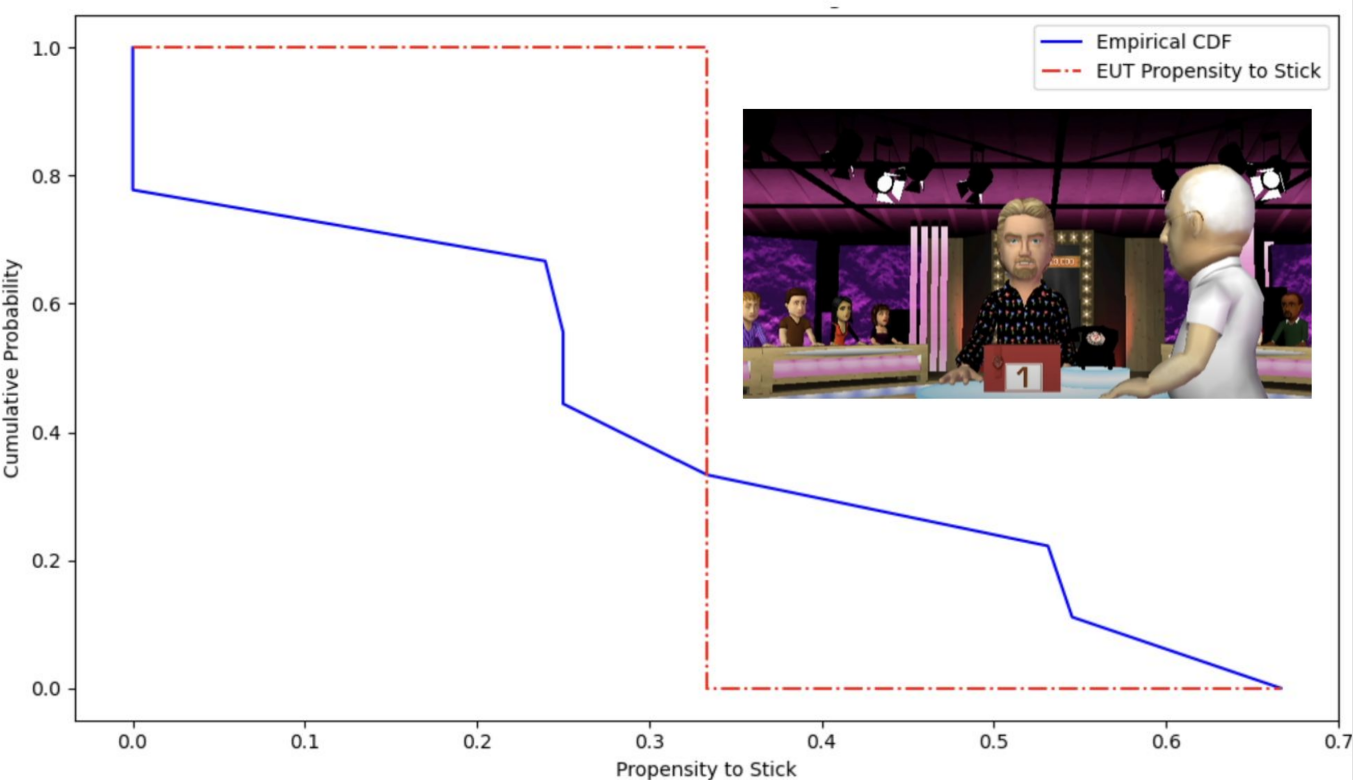}
 \caption{Cumulative probability distribution function for sticking decisions. The inset shows the interface of the video game used in Study~1.\label{Fig1_game}}
\end{figure}

One particular motivation behind these works was the fact that the classical models of cognition and decision-making have been unable to satisfactorily reproduce well-documented behavioural regularities \cite{bardsley2009experimental} and decision making anomalies such as Ellsberg paradox \cite{Ell61} and Machina's re-examination of it \cite{Mac09}. Subsequently, novel models were urgently required. Quantum-mechanical models also help unify and formalise previously heuristically formulated concepts and ideas in the field of psychology \cite{Pot22}. For example, in Ref.~\cite{Kva21} an analogy between varying mental states and oscillating quantum states has been used to explain cognitive dissonance \cite{Fes62}. A similar approach has been adopted in Ref.~\cite{Gro17} to study the phenomenon of confirmation bias \cite{Was60}. Quantum mechanics has also underpinned the models of two-stage gambling, Prisoner's Dilemma game \cite{Pot09} and conjunction fallacy \cite{Bus12, Gro17, Pot22}.

The rationale of applying a quantum theory in cognition and decision-making may be explained comparing the principles of classical and quantum computing. A quantum computer uses a quantum bit (qubit) that can be in the states $|0 \rangle = \left[\begin{array}{@{}c@{}}
    1 \\
    0 
    \end{array} \right]$ and $|1 \rangle = \left[\begin{array}{@{}c@{}}
    0 \\
    1 
    \end{array} \right]$ that are analogous to the `0' and `1' binary states of a classical digital computer. However, a qubit exists in a continuum of states between $|0 \rangle$ and $|1 \rangle$, i.e.~its states can be a superposition $\psi = \alpha |0 \rangle + \beta |1 \rangle$, where the probability amplitudes $\alpha$ and $\beta$ are complex numbers and $|\alpha|^2 + |\beta|^2 = 1$. Quantum algorithms that employ qubits are exponentially faster than any possible deterministic classical algorithm \cite{Nie02}. Subsequently, it has been suggested that models that represent mental states similarly to the possible states of a qubit can better explain certain psychological phenomena that appear paradoxical from a classical perspective \cite{Bus12, Pot22}.
 \begin{figure}
 \includegraphics[width=0.99\textwidth]{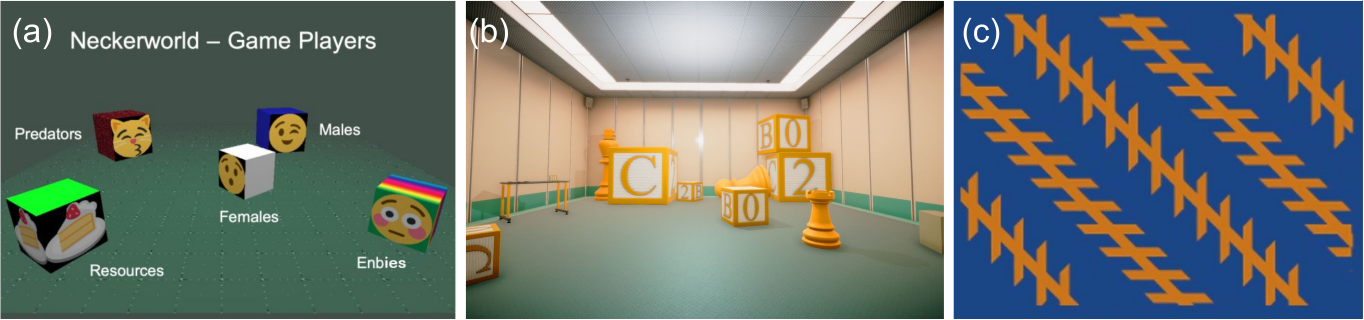}
 \caption{Screenshots exemplifying the use of optical illusions in video games: (a)~Neckerworld---an experimental computer vision game using the Necker cube illusion \cite{Neckerworld}; (b)~Superliminal---a commercial surreal puzzle game that incorporates perspective illusions \cite{game1}; (c)~an experimental video game implementing the Z{\"o}llner illusion \cite{Zol60}, illustrating the potential of visual distortions in gameplay mechanics \cite{wang2021game}.}\label{Fig1_game_examples}
\end{figure}

It should be stressed that the quantum cognition theory \cite{Bus12, Pot22}, which is the mainstream theory of this paper, and the theories collectively known as ``quantum mind hypothesis'' are two distinct approaches to applying principles from quantum mechanics to cognitive science (an extended relevant discussion can be found in Ref.~\cite{Mak24_APL}). The quantum cognition theory uses the mathematical framework of quantum mechanics as a metaphorical tool to model cognitive phenomena, particularly decision-making and perception. Rather than suggesting that the brain or mind operates according to actual quantum physical processes, quantum cognition treats cognitive states, choices and perceptions as probabilistic, using quantum probability theory to capture the complex, sometimes contradictory, patterns of human reasoning and perception \cite{Khr06, Bus12, Pot22}. For example, quantum cognition models explain how individuals may hold superposed cognitive states or shift between incompatible beliefs when confronted with new information \cite{Bus12, Mak24_information, Mak24_information1}. Importantly, quantum cognition does not imply that the brain itself operates as a quantum computer. Instead, it leverages the probabilistic nature and superposition principles of quantum mechanics to explain mental processes that classical probability struggles to describe. In contrast, quantum mind theories propose that consciousness and cognitive processing may indeed be underpinned by quantum physical processes occurring in the brain. These theories posit that genuine quantum states exist in biological systems, potentially at the level of neurons, where subatomic quantum effects such as entanglement or superposition might influence cognition directly \cite{Ham96, Koch06, Geo18, Geo20, Geo21, Geo22, Geo_book}. While these theories are not directly applied in this paper, the concepts underlying them often complement the methods used in the quantum cognition theory and, therefore, should not be entirely disregarded.

In this paper, we show the value of quantum cognition modelling using 3~examples. In Study~1, we use the data from ``Deal or No Deal'' video game decision-making experiment to demonstrate the need for quantum modelling of the underlying decision processes. In Study~2 of the current work, we address the problems of discrepancy between experimental data and the traditional representation of perceptual reversal as a series of abrupt switches between two discrete states illustrating the underlying behavioural phenomena using examples of several video games with optical illusions. In Study~3 we generalise and numerically validate a physiology-consistent quantum-physical model of bistable perception based on the Schr{\"o}dinger equation. Finally, we discuss the ability of our model to capture diverse optical illusions and other psychological effects such as cognitive dissonance.

\section*{Study~1: The Need for New Quantum Approach to Understand Human Behaviour}
The video game ``Deal or No Deal UK'' offers a controlled yet dynamic environment to test human decision-making under risk, illustrating the need for quantum-inspired approaches to model cognitive behaviours (Fig.~\ref{Fig1_game}). In the TV show version of the game, players receive an opportunity to swap their box for any of the remaining boxes in the game only once during the show, when there are two unopened boxes left. Yet, the decision to swap the box or to stick to the original box represents a unique opportunity to test whether the participants behave according to Expected Utility Theory (EUT) or to Cumulative Prospect Theory (CPT).

Using the data from the TV shows broadcast in France, Italy, and the UK, Blavatskyy and Pogrebna \cite{blavatskyy2010endowment} have previously demonstrated that: (i)~CPT players should be more likely to stick to their original choice of the box due to the embedded assumption of loss aversion in the Cumulative Prospect Theory and (ii)~EUT players should be exactly indifferent between swapping and sticking. While it has been demonstrated that the TV show contestants are more likely to behave according to the EUT than CPT theory, the TV show data on swapping and sticking choices are limited due to the fact that many players accept the banker's offer early and, therefore, do not get to the last round of the game, where there are only two unopened boxes left. 

At the same time, a video game setting allows exploring the possibility to swap boxes more than once. Taking advantage of this opportunity, we recruited 78 study participants (members of the general public over 18 years of age) to participate in the decision making experiment involving the ``Deal or No Deal UK'' video game platform. All experimental procedures adhered to relevant ethical standards and guidelines, with appropriate approval obtained from the Humanities and Social Sciences Research Ethics Committee of the University of Warwick. Participants provided informed consent before participating in the study. The recruited players made a total of 1,698 decisions, of which 486 were swap or stick decisions (i.e.,~they were asked to follow the video game instructions as usual, but were offered to swap their boxes after they received each of the bank offers by the experimenter).

{\Ivan The study involved 78 participants (43 women and 35 men) ranging in age from 19 to 45 years (M = 28.3, SD = 5.9). Participants were recruited via university mailing lists, research participant pools and online advertisements. Each participant received a fixed payment of \pounds15 for their involvement, with additional performance-based incentives tied to their decisions in the game. The experiment was conducted in a controlled laboratory environment at the University of Warwick, where participants completed the tasks individually on computer stations to maintain decision-making independence. Each session lasted approximately 60 minutes, encompassing the consent procedure, task instructions, experimental trials and debriefing.}

Interestingly, all 78 players played the game until the end, so that each player received 8 offers from the banker. Despite the fact that 63\% of decisions were stick decisions and 37\% of decisions were swap decisions, demonstrating the overall consistency with EUT, a closer look at the Cumulative Distribution Function (CDF) of swap or stick decisions demonstrates an intriguing result: while the EUT prediction may imply an approximately equal split between swapping or sticking decisions, the actual functional form of the distribution of swap or stick decisions observed in the video game experiment is not a step-wise but a continuous function, not precisely captured by the EUT prediction (see the inset in Fig.~\ref{Fig1_game}).

This finding highlights the need for a more sophisticated model of decision making. Indeed, looking at the results presented in the inset in Fig.~\ref{Fig1_game} from a point view of physics, engineering, chemistry or biology, the behaviour of the empirical data curve (the solid line) immediately resembles the response of a dynamical system \cite{Str15, Mak24_information1}. While dynamical systems can be studied using methods of classical physics in principle \cite{Mak24_information1}, a comprehensive understanding of their properties often requires applying the methods of quantum mechanics \cite{Leo90, Pan05, Wim22, Mak24_information}, which is the fact that shapes the research approach adopted in this present paper.  

\section*{Quantum-Mechanical Model and Bistable Perception}
As discussed above, a more sophisticated and precise model capable of going beyond the EUT and the CPT predictions could be devised based on the quantum-mechanical principles \cite{Gri04}. Furthermore, quantum-inspired models could be showcased through the analysis of optical illusions that are often observed in video games, including Superliminal \cite{game1} and Neckerworld \cite{Neckerworld} (Fig.~\ref{Fig1_game_examples}) as well as Monument Valley, Anamorphosis, Perfect Angle and I Spy Universe, since, in particular, such models underline the significance of considering superposition states in explaining human perception \cite{wang2021game, Mak24_APL}. 

\begin{figure}
 \includegraphics[width=0.79\textwidth]{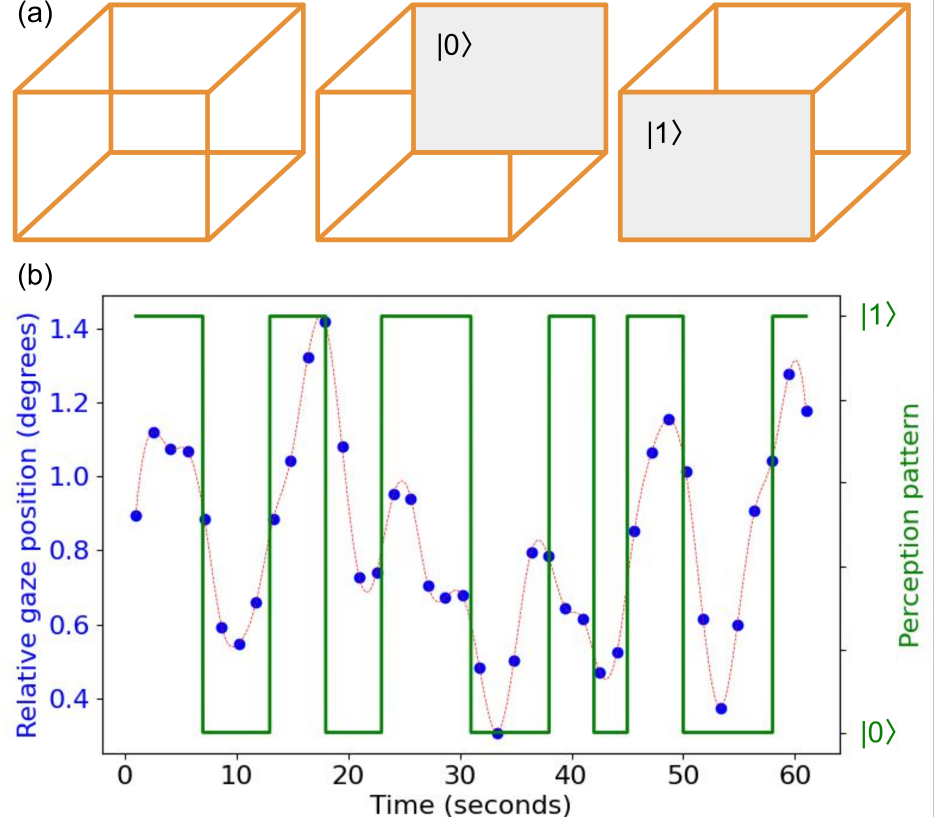}
 \caption{(a)~The Necker cube and its two possible stable interpretations denoted as the $|0 \rangle$ and $|1 \rangle$ states in the main text. {\Ivan(b)~Typical experimental discrete perception pattern (green solid line, right $y$-axis), where the cube can be either in $|0 \rangle$ or $|1 \rangle$ state \cite{Cho20}. The continuous dashed line is the guide to the eye between the experimental data points (blue dots, left $y$-axis) corresponding to an eye-tracking signal measured simultaneously with the state of the Necker cube reported by an observer \cite{Cho20}.}\label{Fig1}}
\end{figure}

This strategy has been applied in Ref.~\cite{Bus12} to model the bistable perceptual processing of the Necker cube (Fig.~\ref{Fig1}a), which is an optical illusion caused by an ambiguous graphical representation \cite{Nek32, Lon04} and is showcased in the Necker board game (see Ref.~\cite{game2} for more detail). In contrast to a classical Markov model of bistable perception, where the possible states of the cube were `0' and `1' \cite{Bus12}, in the quantum model the states of the cube can be in a superposition of the states $|0 \rangle$ and $|1 \rangle$. When the perception state is measured, the quantum state corresponding to it collapses from a superposition of $|0 \rangle$ and $|1 \rangle$ to one of the stable perception states of the cube (Fig.~\ref{Fig1}a). Hence, the quantum model can account for multiple outcomes while processing input data with a large set of constraints. The outputs of the quantum model are probabilistic and they have been shown to describe human mental states more efficiently than classical models of similar complexity \cite{Bus12}.

The implementation of qubits in an experimental physical system (e.g.~electrons) could be used to model the human decision-making and perception, also having implications for our understanding of complex cognitive processes, should the still-debated quantum mind hypothesis gain wider acceptance (e.g.~Refs.~\cite{Khr06, wendt2015quantum, mindell2012quantum}). However, while such physical experiments are feasible, quantum measurements are technically challenging. On the other hand, quantum-physical systems can be studied using the well-known mathematical models such as the Schr{\"o}dinger equation. This approach is adopted in this present work to demonstrate that a mathematical model of a real-life physical system could be used to bridge a gap between theoretical and experimental results obtained in the studies of psychological phenomena.

The Schr{\"o}dinger equation is a partial differential equation that governs the wave function of a quantum-mechanical system \cite{Gri04}. For a single electron that exists in a one-dimensional space it can be written as
\begin{equation}
  \label{eq:SE}
  i\hbar\frac{\partial \psi(x,t)}{\partial t}=\left[-\frac{\hbar^2}{2m}\frac{\partial^2}{\partial x^2} + V(x)\right]\psi(x, t)\,, 
\end{equation}
where $\psi(x, t)$ is a wave function, $i$ is the imaginary unit, $m$ is the mass of the electron, $\hbar$ is Plank's constant and $V(x)$ is the potential that represents the environment where the electron exists. 

In the work Ref.~\cite{Bus12}, the space and time variables of Eq.~(\ref{eq:SE}) were algebraically separated, resulting in an equation that predicts the existence of stationary states. Interpreting such states as probabilities of perception of the states of the Necker cube, it was demonstrated that the probability harmonically alternates from zero to one as a square sine function of time. A similar result was obtained in Ref.~\cite{Atm10}, where some principles of quantum mechanics were used to describe the switching of mental states during the perception of ambiguous figures.

{\Ivan However, the theoretical results produced by the aforementioned quantum models cannot fully explain results obtained in neuroscience experiments involving the perception of the Necker cube and other ambiguous figures. For example, in Refs.~\cite{Run16, Pia17, Joo20}, electroencephalograms were recorded consistently with subjective inputs given by observers. Whereas those inputs were represented by abruptly changing square pulses of a seemingly random duration, the envelopes of the corresponding experimental data were continuously changing pulse-like waveforms. Similar continuously varying waveforms were observed in eye-tracking experiments (Fig.~\ref{Fig1}b in this paper; see Fig.~1b in Ref.~\cite{Cho20} for the original data), where a blink or movement of eye could be associated with a perceptual reversal \cite{Lon04, Ang20}. Complementary experimental evidence has also been produced by the studies relating perceptual decisions with eye movements \cite{Mat23}. Nevertheless, we emphasise that any comparison with experimental data in this work is made solely at a reference level, as the primary objective of this study is to introduce a conceptually new theoretical model. If required, a stronger connection to experimental data can be established, as evidenced by the successful replication of experimental results using artificial neural networks specifically designed to mimic the perception of optical illusions \cite{Ara20, Mak24_APL}.}

We also comment on a fundamental connection between the behavioural pattern observed in Study~1 (Fig.~\ref{Fig1_game}) and the one depicted in Fig.~\ref{Fig1}b and investigated in the remainder of this paper. While the former pattern is plotted as a function of human preference but the latter one is a function of human perception that varies in time, both reflect the outcome of a reversal process (i.e.~switching between two alternative states). A connection between perceptual and preferential decision-making has been known from the psychological point of view \cite{Dut16}. A similar connection can also be established considering this problem from the physical perspective.

In physics, a bistable switching system can be encountered in either of two possible states \cite{Luo13}. For example, a mechanical light switch can be encountered either the `ON' or `OFF' position. Similarly, bistable electronic, magnetic and optical devices can store one bit of data, with one state being `0' and the other state being `1'.

The light switch cannot operate when it is positioned between the `ON' and `OFF' states, which enables us to draw an analogy between its behaviour and the step-wise curve in Fig.~\ref{Fig1_game}. However, in electronic, magnetic and optical devices the switching between the `0' and `1' states is not instantaneous but rather gradual with physically possible intermediate states. The physical processes behind such a switching mechanisms can be investigated both as a function of time and as a function of current density or light intensity applied to the particular device to control the switching \cite{Mak12, Mon19}. The latter scenario can be considered to be a conceptual counterpart of the continuous empirical data curve in Fig.~\ref{Fig1_game}. It is noteworthy that a rigorous description of the physical processes underlying the operation of electronic, optical and magnetic devices requires applying the methods of quantum mechanics. Indeed, for example, the phenomenon of magnetism is intrinsically quantum-mechanical in nature and magnetic ordering can only be explained using a quantum theory \cite{Ste13}.  

\section*{Results}
\subsection*{The model}
\begin{figure}
 \includegraphics[width=0.89\textwidth]{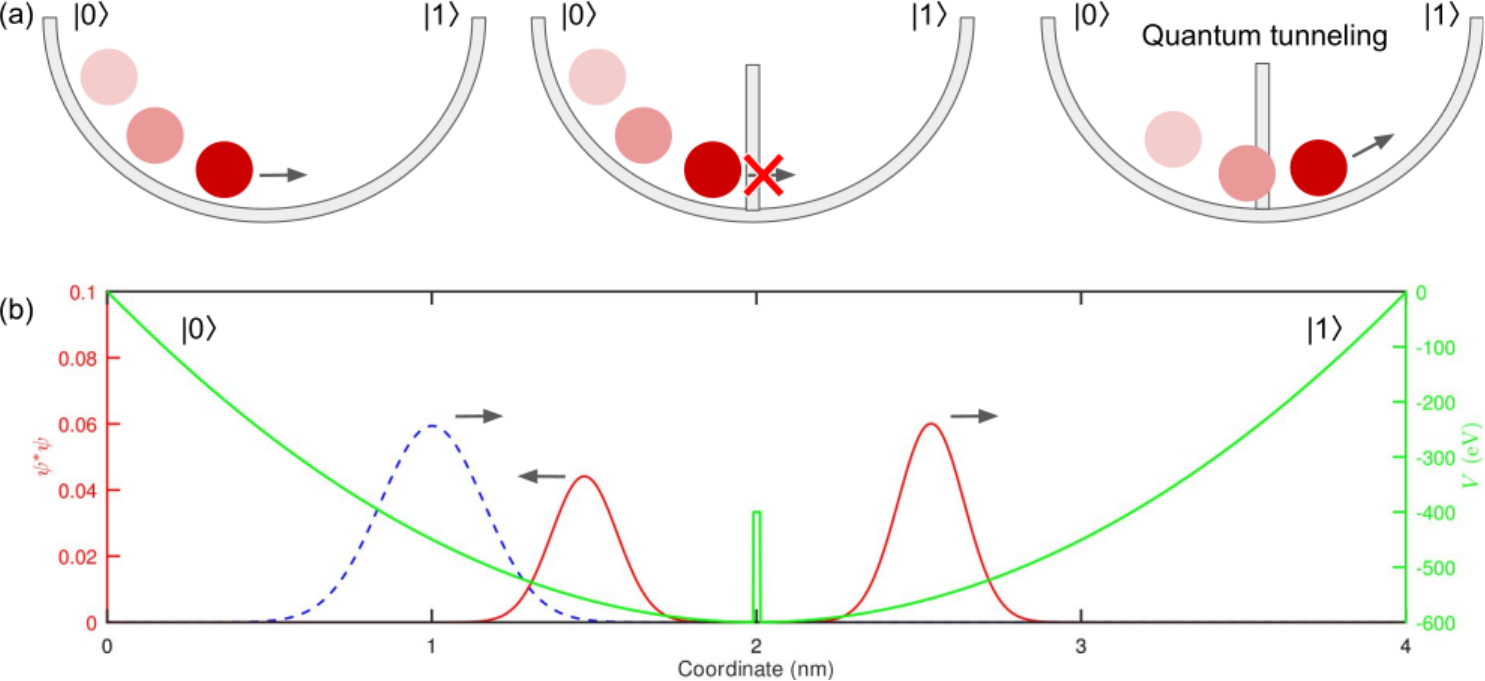}
 \caption{(a)~In classical mechanics, a ball that typically rolls back and forth (i.e.~undergoes harmonic motion) inside an empty bowl cannot surmount a barrier placed on its way. In quantum mechanics, an electron trapped in a parabolic well behaves as a harmonic oscillator and it can tunnel (pass) through a barrier. (b)~Result of a numerical simulation of quantum tunnelling through the barrier. The labels $|0\rangle$ and $|1\rangle$ correspond to the states of the Necker cube.\label{ball_in_the_bowl}}
\end{figure}

We analyse a physical system consisting of an electron that undergoes a harmonic motion in a parabolic potential well (Fig.~\ref{ball_in_the_bowl}a) to capture the human behaviour reversals and perceptional reversals that can be illustrated using examples of ``Deal or No Deal'' video game as well as the computer games that employ optical illusions (e.g.~Superliminal, Monument Valley, Anamorphosis, Perfect Angle and I Spy Universe). In classical mechanics, a counterpart of this system would be a small idealised ball that constantly rolls back and forth inside a bowl. Whereas the ball may not have enough energy to surmount or penetrate a physical barrier placed inside the bowl, in a similar scenario the electron may pass through the barrier by virtue of the quantum tunnelling effect (Fig.~\ref{ball_in_the_bowl}a).

In Fig.~\ref{ball_in_the_bowl}b we present the results of modelling of the electron tunnelling through a potential barrier located in the middle of a parabolic potential well. The $x$-axis is the spatial coordinate across the potential well. The right $y$-axis plots the $V(x)$ profile of the barrier. The left $y$-axis plots the modulus square of the wave function $\psi$ obtained by solving Eq.~(\ref{eq:SE}). The wave function $\psi$ has the meaning of probability amplitude but its modulus square corresponds to the probability density of finding the electron at a certain position in space.

Representing the electron as a Guassian-shape energy wave packet that moves towards the potential barrier (the blue dashed curve in Fig.~\ref{ball_in_the_bowl}b), we observe that one part of the incident wave packet becomes reflected from the barrier but another part is transmitted through it. This result means that there is a certain probability that the electron passed through the barrier and a certain probability that it has been reflected from the barrier.

We virtually split the parabolic potential well into two equal spatial regions and we denote them as $|0\rangle$ and $|1\rangle$. We compute the probabilities of finding the electron inside those regions and we interpret them as the probabilities of perceiving the Necker cube in the states $|0\rangle$ and $|1\rangle$, respectively. In Fig.~\ref{ball_in_the_bowl}b the electron starts moving from the left side of the potential well, which means that, at the beginning of the simulation, the probability of finding the electron in the region $|0\rangle$ is $P_{|0\rangle}=1$ (note that $P_{|0\rangle}+P_{|1\rangle}=1$). At the end of the simulation, analysing the wave packets that have been reflected from and transmitted through the barrier, we obtain $P_{|0\rangle}=0.35$ and $P_{|1\rangle}=0.65$.

\subsection*{Study~2: Bistable perception}
\begin{figure}
 \includegraphics[width=0.89\textwidth]{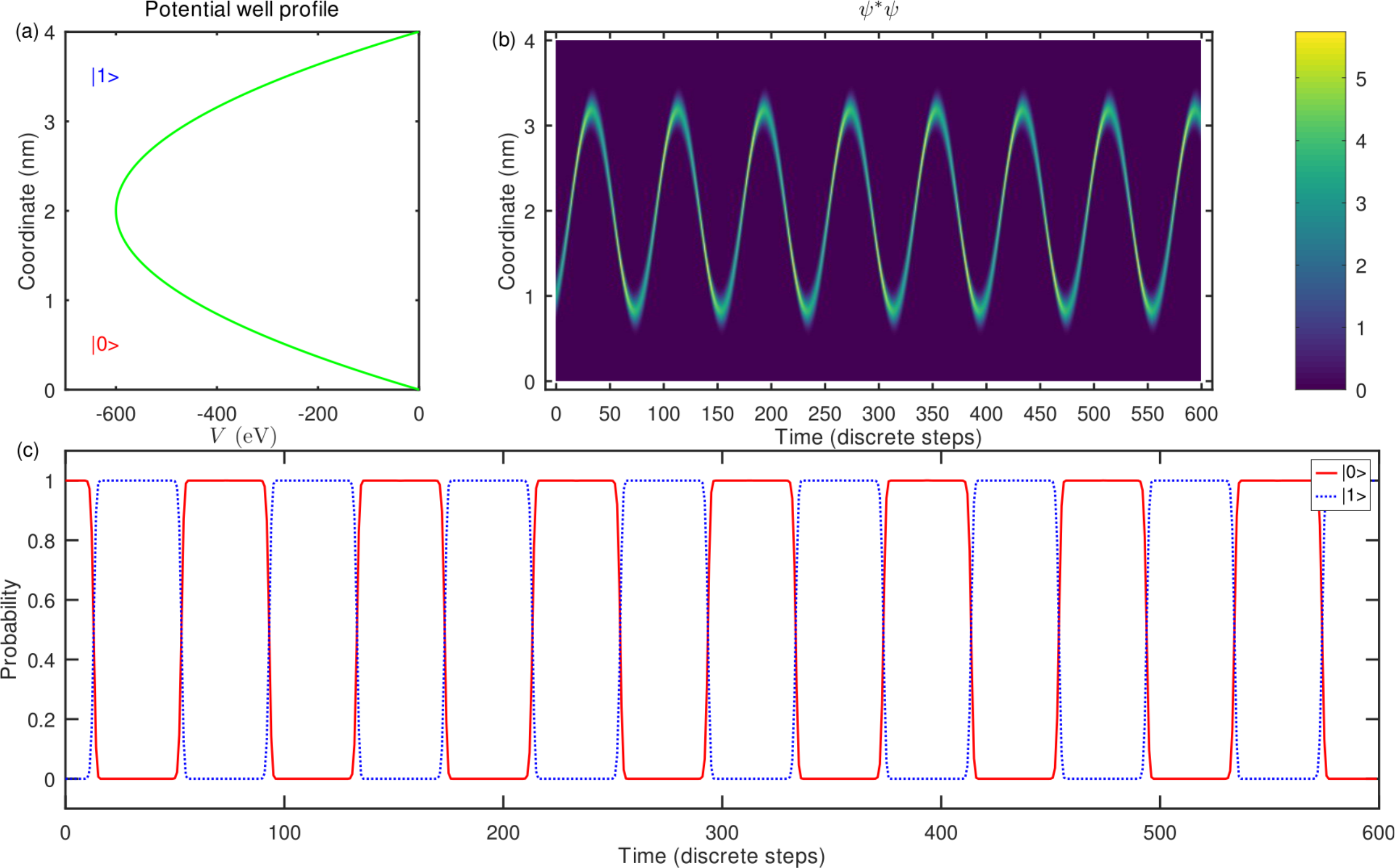}
 \caption{(a)~Model using a potential well with a single parabolic profile. The electron wave packet originates at approximately 1\,nm and it moves in the positive direction. (b)~False colour map plot of the probability density $\psi^{*}\psi$ plotted as a function of both the spatial coordinate inside the potential well and the discrete time. (c)~Probability of finding the electron in the states $|0 \rangle$ (solid curve) and $|1 \rangle$ (dotted curve) as a function of discrete time.\label{parabolic}}
\end{figure}

We start with the discussion of a single parabolic potential well (Fig.~\ref{parabolic}a). We plot the numerical data as a false-colour map (Fig.~\ref{parabolic}b), where the blue colour corresponds to the zero probability density but the yellow colour denotes the maximum value of the probability density. One can understand this colour map as a set of individual plots, such as the one in Fig.~\ref{ball_in_the_bowl}b, that are arranged together in a chronological order.

We can observe a harmonic oscillator behaviour, as a result of which the probability of finding the electron in the $|0\rangle$ and $|1\rangle$ states periodically varies with time (Fig.~\ref{parabolic}c). While the electron is more likely to be found in the pure $|0\rangle$ and $|1\rangle$ states, its transition between them is not instantaneous but gradual. This means that the electron exhibits a temporal nonlocality since its states are not sharply localised but continuously change over a time interval \cite{Atm10}. For example, at certain instances of time, the probability of finding the electron in the states $|0\rangle$ or $|1\rangle$ equals 0.5. This situation may also be interpreted as a temporally interleaved percept-choice sequence known in the context of neuro-dynamical bistable perception models \cite{Noe12}. Importantly, the result in Fig.~\ref{parabolic}c is in good agreement with the simulated data produced using the quantum-mechanical models in Ref.~\cite{Atm10} and Ref.~\cite{Bus12}.

The choice of the origin of the electron wave packet in Fig.~\ref{parabolic} deserves additional discussion. Strictly speaking, the origin of the wave packet should be chosen according to the Necker cube orientation preferred by the observer. However, every observer has their own preferred initial orientation and that orientation can change from one experiment to another. To account for this effect, the coordinate of the origin may be chosen randomly. However, the realisation of the this scenario is computationally demanding and involves additional complex post-processing steps. Thus, for the sake of the discussion of the main results, in the following we will continue assuming that the wave packet originates at the coordinate of approximately 1\,nm and that it moves in the positive coordinate direction. The choice of other origin coordinates leads to qualitatively similar results and, therefore, it is inconsequential for the overall discussion.

\begin{figure}
 \includegraphics[width=0.89\textwidth]{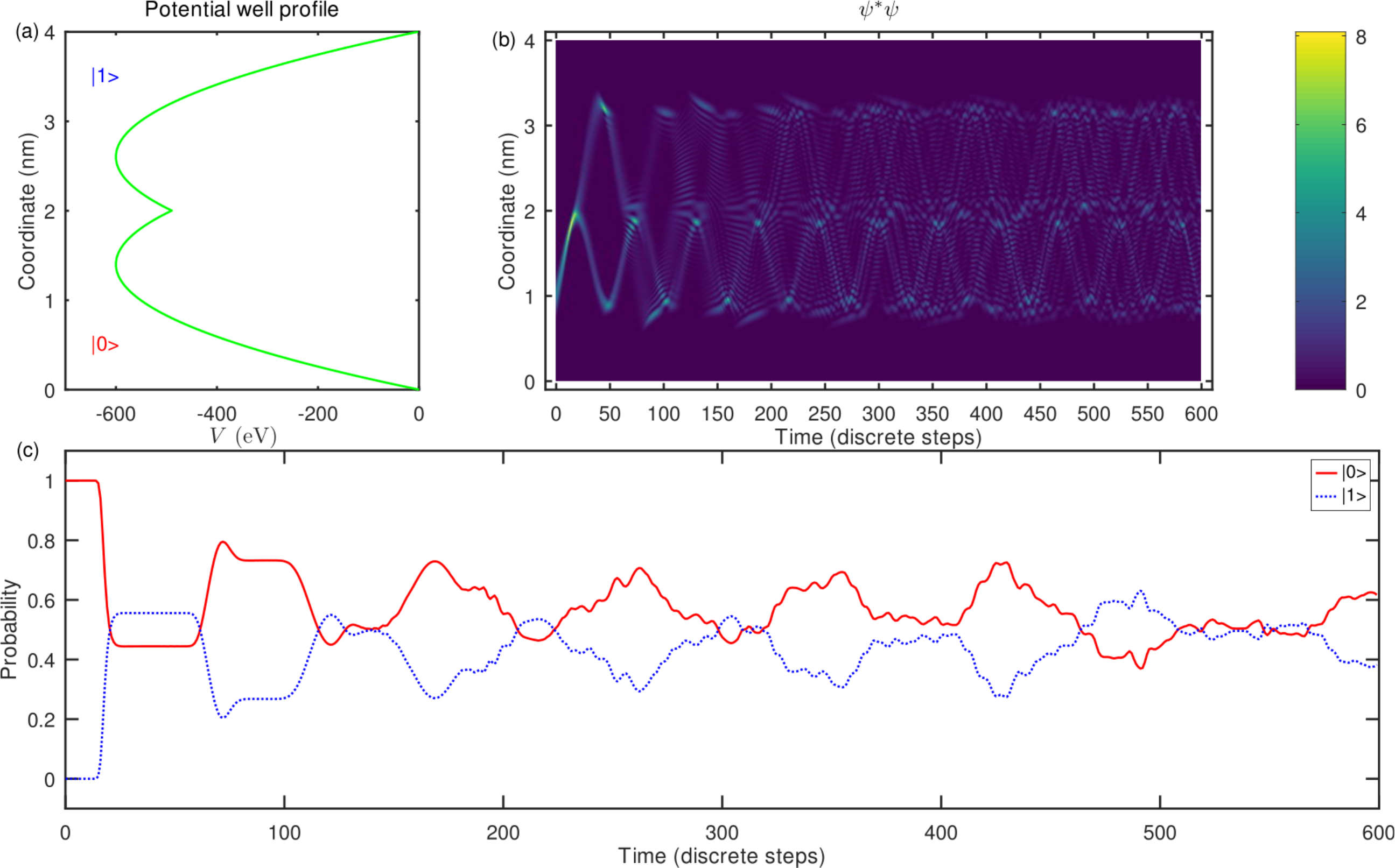}
 \caption{(a)~Model using a potential well with a symmetric double parabolic profile. (b)~False colour map plot of the probability density $\psi^{*}\psi$ plotted as a function of both the spatial coordinate inside the potential well and the discrete time. (c)~Probability of finding the electron in the states $|0 \rangle$ (solid curve) and $|1 \rangle$ (dotted curve) as a function of discrete time.\label{double_parabolic}}
\end{figure}

However, the predictions of the single parabolic potential well model do not fully agree with the experimental data obtained in the tests involving the Necker cube (see Fig.~\ref{Fig1}b). To resolve this problem, we use a double parabolic well with a barrier located in the middle of the computational domain (Fig.~\ref{double_parabolic}a). In this updated model, the electron undergoes quantum tunnelling through the barrier, which gives rise to wave function interference processes that are analogous to interference effects observed in the fundamental Young's slit interference experiment \cite{Jak64, Suz08, Guo21}.

Our double-potential well model is inspired by the previous models developed to simulate bistable perception of the Necker cube and other ambiguous objects, where, in particular, the two principal perceptual states of the Necker cube were phenomenologically represented by a two-level quantum-like systems \cite{Mil09, Con15, Ben18, Atw14}. Relevant double-potential well models that combine certain concepts of classical physics and statistics have also been put forward in Refs.~\cite{Pas12, Run16, Mei19, Lep20}, where nonuniform energy wells were used to represent reversals of perception. However, none of the those models relies on the analysis of the Schr{\"o}dinger equation, though a relevant idea was put forward in Ref.~\cite{Ben18}, including a suggestion that quantum tunnelling might play a role in bistable perception.

It is also noteworthy that in the aforementioned models the profiles of energy wells and the height of the energy barriers between them have been typically chosen based on phenomenological observations of bistable perception of ambiguous figures. Since the probability of tunnelling through the barrier decreases as the barrier height is increased, a meaningful numerical simulation always implies a judicious choice of the barrier height based on available experimental data and expertise of the developer of the model.      

In Fig.~\ref{double_parabolic}b, where we present the results of the simulation of a symmetric double parabolic profile, the energy wave packet originates from the well with a lower spatial coordinate value and it moves in the positive direction. We can see that at the discrete time of approximately $T=25$ the wave packet splits into two approximately equal waveforms due to the interaction with the barrier.

The corresponding probabilities of finding the electron in the states $|0 \rangle$ and $|1 \rangle$ are plotted in Fig.~\ref{double_parabolic}c as a function of time, where we can observe a periodic switching between the fundamental perception states. However, apart from a short period at the beginning of the simulation where the system is temporarily strongly affected by the input, we can see that the electron is mostly found in a superposition of the $|0 \rangle$ and $|1 \rangle$ states due to wave function interference processes. Such a superposition has often been attributed to the potential existence of mental ``superposition'' states \cite{Atm10, Bus12}. 

The result in Fig.~\ref{double_parabolic}c also aligns with the outcomes of a recent test of perceptual reversals of the Necker cube \cite{Wil23}, where it has been shown that perception can become unstable before the actual reversal event is reported by the viewer. Yet, observers may intentionally control their perception of the cube, thus changing the number of perceptual reversals over time \cite{Lon04}. However, the reversal cannot be prevented entirely. The result in Fig.~\ref{double_parabolic}c reproduces both varying number and time duration of perceptual reversals, also confirming the impossibility of a complete prevention of the reversal. 

Perceptual ambiguity does not mean that two alternative interpretations are always equally possible. Instead, there is typically a preferred interpretation that an observer tends to see \cite{Nak11}. Moreover, the perceptual reversal rate often depends on the familiarity of the ambiguous figure, adaptation and concentration \cite{Lon04, Sto12}. 

\begin{figure}
 \includegraphics[width=0.89\textwidth]{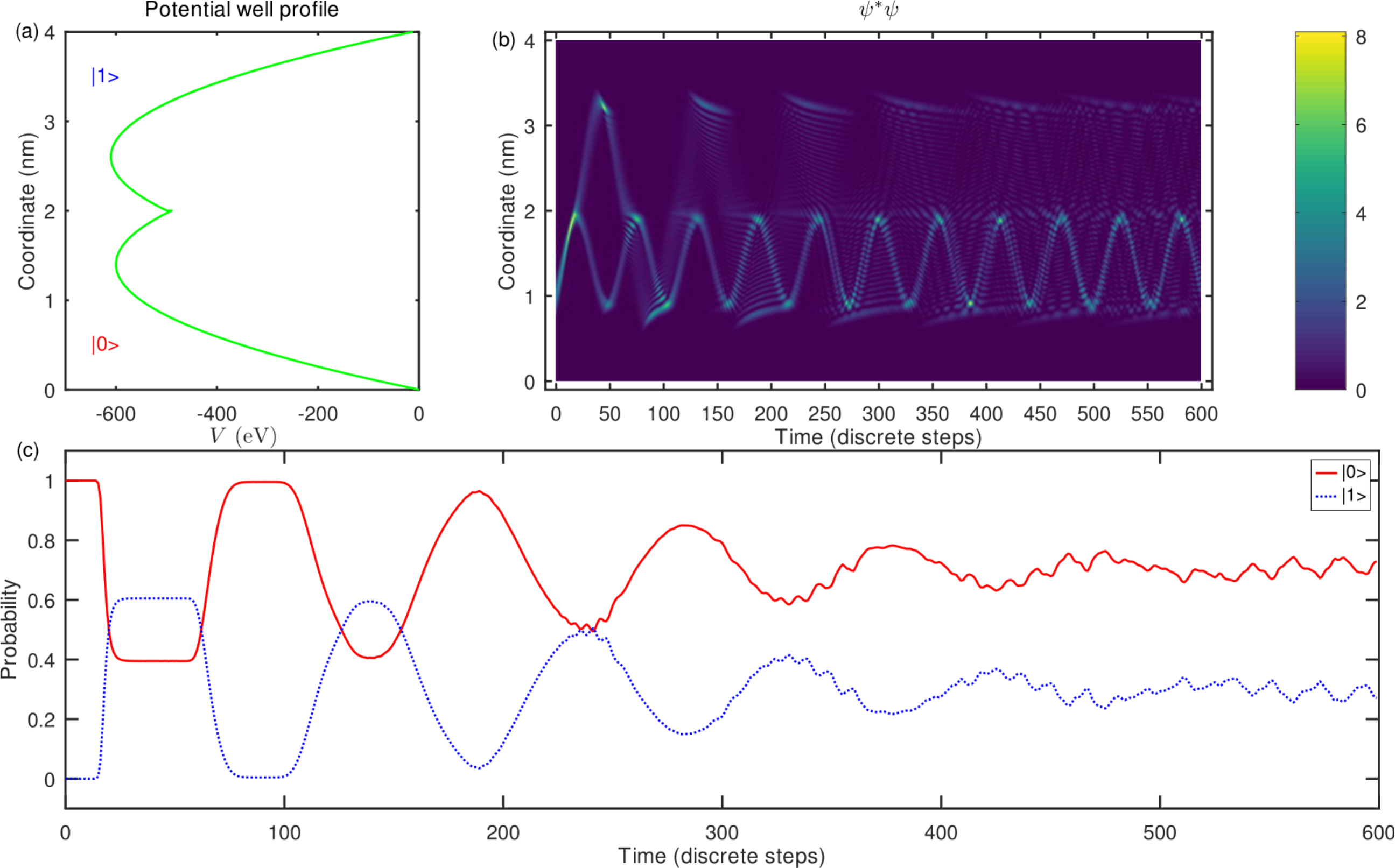}
 \caption{(a)~Model using a potential well with an asymmetric double parabolic profile obtained by shifting one of the wells by $-10$\,eV. (b)~False colour map plot of the probability density $\psi^{*}\psi$ plotted as a function of both the spatial coordinate inside the potential well and the discrete time. (c)~Probability of finding the electron in the states $|0 \rangle$ (solid curve) and $|1 \rangle$ (dotted curve) as a function of discrete time.\label{asym_double_parabolic}}
\end{figure}

Our model enables us to adjust the behaviour of the electron to account for these processes. In Fig.~\ref{asym_double_parabolic}a, we use an asymmetric double parabolic well with a higher barrier compared with that in Fig.~\ref{double_parabolic}a. The probability of electron tunnelling through this barrier decreases accordingly, thereby increasing the probability of finding the electron inside the same well from where it originates (Fig.~\ref{asym_double_parabolic}b,~c). We interpret this result as a preference that can be induced by changing the gaze direction and suppressing eye blinking \cite{Lon04}. The model also shows that perceptual reversals cannot be suppressed because it is impossible to hold a preferred cube orientation for a long time.

\subsection*{Study~3: Link between physics and physiology of bistable perception}
While previous studies have successfully employed overarching principles of quantum mechanics to model the perception of ambiguous figures, a comprehensive link between the physics and physiology of bistable perception has remained elusive. The present paper endeavours to bridge this gap, establishing a concrete connection between these two realms. We also suggest that the physical processes that underpin the model proposed in this work might correspond to an eye blink, which is an action that leads to a change in the perception state \cite{Lon04}. Moreover, the eye blink has been associated with the ability to better recognise objects \cite{Ang20}. Subsequently, the analysis conducted in this section should contribute to the studies of enhanced visual perceptual functions of action video game players compared with non-video game playing individuals \cite{Poh22}. Therapeutic treatments that improve perceptual skills via action video game training \cite{Li15} may also benefit from our findings. 

Quantum harmonic oscillations result in a continuous conversion of energy from kinetic to potential and vice versa. Yet, due to the potential barrier located between the adjacent parabolic wells, the energy conversion processes are accompanied by redistribution of energy from a region where destructive wave function interference occurs to a region of constructive interference.

Although the human eye is often considered to be sensitive only to the intensity of light, it has been argued that optical illusions and cognition may be caused by phase-sensitive optical wave phenomena \cite{Gla15}. Variations of phase are also important in the context of quantum interference, which, in the framework of our model, means that phase plays a role in defining the probability of finding the electron in the states $|0\rangle$ and $|1\rangle$ that correspond to the perceptual states of the Necker cube. 

A blink is important for maintaining the lubrication of the eye: a thin layer of fluid that covers the outer mucosal surfaces of the eye is renewed every time the eye blinks \cite{Sha14}. From the point of view of optical wave physics, the formation of this liquid film inevitably leads to optical interference processes and phase shifts \cite{Gru05, Tem15}.

While, to our best knowledge, any connection between the optical properties of the lubricated eye and bistable perception has not been demonstrated, several plausible arguments can be provided in favour of it. For example, it is known that aged people often suffer from the syndrome of dry eyes and from reduced amplitude and speed of eye blinks \cite{Sha14}. At the same time, many experiments have revealed a lower ability of aged people to experience bistable stimuli \cite{Pat16}. Subsequently, a connection between wave-optical changes in the eye due to eye blinking and lubrication and bistable perception cannot be ruled out.

\begin{figure}
 \includegraphics[width=0.89\textwidth]{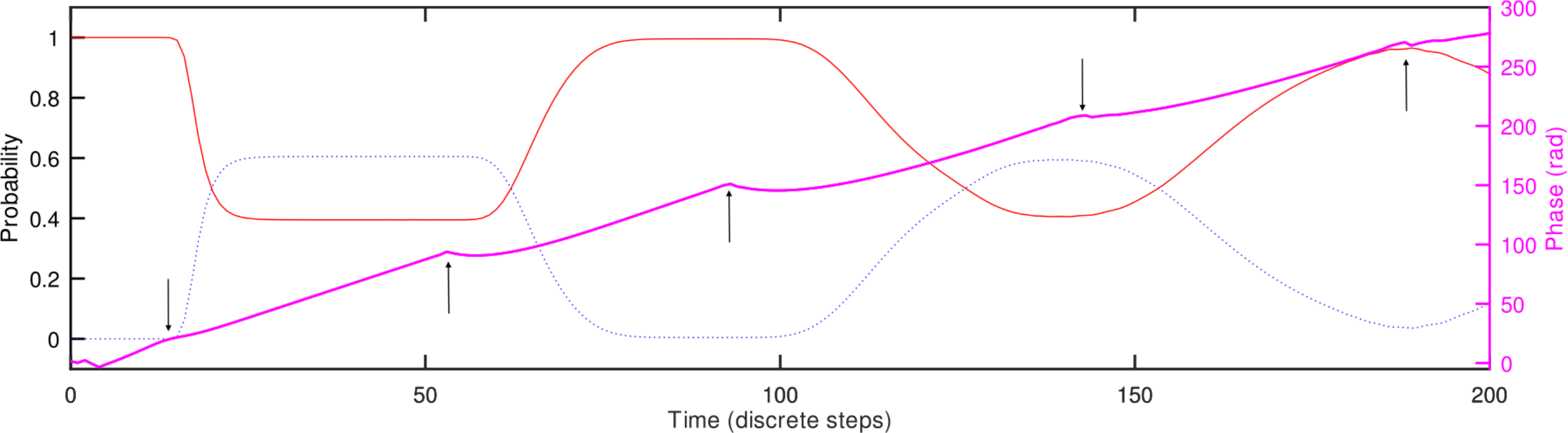}
 \caption{Probability of finding the electron in the states $|0 \rangle$ (solid curve) and $|1 \rangle$ (dotted curve) in the asymmetric double parabolic well plotted alongside the phase shift. The arrows highlight the changes in phase that herald the changes in the perception state.\label{phase}}
\end{figure}

Thus, we suggest that in our model a phase shift should be associated with eye blinks. In Fig.~\ref{phase}, we plot the probability of finding the electron in the states $|0 \rangle$ and $|1 \rangle$ (the same data as in Fig.~\ref{asym_double_parabolic}c but plotted on a shorter time scale) alongside the phase detected in the middle of the potential barrier formed by the adjacent parabolic wells. We reveal that every perceptual reversal is heralded by shifts in an otherwise monotonically increasing phase of the wave function. Interestingly, this results qualitatively agrees with the experimental observation of perception destabilisation that precedes the perceptual reversal by approximately one second \cite{Wil23}. Yet, the phase shifts occur closer to the end of the stable perception periods and they are followed by a gradual change in the probability. A similar behaviour was predicted previously in \cite{Bus12} and it is consistent with the fact that an eye blink improves the ability to recognise objects. Naturally, the changes caused by an eye blink cannot occur instantaneously, which explains, for example, a gradual decrease in the probability to perceive the state $|0 \rangle$ at the discrete time of approximately $T=100$ in Fig.~\ref{phase}.    

\section*{Discussion}
\subsection*{General discussion}
Our quantum-mechanical model offers unique insights into cognitive processes observed in video games, including bistable perception and decision-making under uncertainty. By integrating quantum principles into game mechanics, developers could design games that probe cognitive dynamics, offering entertainment, learning, therapy, as well as a deeper understanding of human behaviour. We tested several ambiguous two-dimensional drawings, including Rubin's vase, ``My Wife and My Mother-in-Law'', rabbit-duck illusion and Spinning Dancer illusion, the principles of which are used in many video games mentioned above, and we established that changes in their perception are essentially similar to those of the Necker cube but follow different temporal dynamics. Our observations agree with the previously published data that reveal a high correlation between the switching rates of different bistable stimuli \cite{Cao18}.

Importantly, being based on a numerical model of a realistic physical process, our approach is universal in terms of its ability to capture perceptual and behaviour reversals that either occur as a function of time or triggered by human decision-making processes. The universality of the model stems from the fundamental properties of dynamical systems and the ability to represent their evolution both in time domain and phase space \cite{Str15}.   

Our model can capture the differences in temporal dynamics using different profiles of the potential well. For example, longer switching times can be achieved using a quartic oscillator and different types of double potential wells with a Gaussian perturbation \cite{Gri04}. Optimisation and machine learning techniques can also be used to produce a unique profile of the potential well that would reproduce experimental data. More complex two-dimensional potential wells can also be used. Moreover, the model can potentially be generalised using an $N$-dimensional Schr{\"o}dinger equation \cite{Iny22}, thus enabling, for example, the modelling of multistable perceptual states \cite{Bus12}. One can also model psychological phenomena using periodic distributions of potential wells \cite{Mak24_information}.

It can also be demonstrated that the same model can be applied to cognitive dissonance, a mental discomfort experienced by a person, who holds two or more contradictory beliefs, ideas or values. Previous works showed that temporal oscillations serve as a plausible model of cognitive dissonance \cite{Kva21}. Independently, it has been experimentally demonstrated that the oscillatory activities in the brain play an important role in bistable perception and other complex cognitive processes \cite{Sen20}. 

We suggest that a quantum harmonic oscillator system can be used to simulate oscillations of mental states between two or more contrasting beliefs and ideas. In such a model, the contradictory beliefs and values could be assigned specific priorities using potential wells of arbitrary complex configuration. Moreover, an energy barrier and the quantum tunnelling through the barrier could represent a psychological barrier that needs to be overcome to resolve a conflicting situation \cite{Fri19}. 

We note that the model of cognitive dissonance proposed in Ref.~\cite{Kva21} is effectively an extension of the classical and quantum-mechanical models of bistable perception previously used in Ref.~\cite{Bus12}. Since our quantum harmonic oscillator model has been inspired by the models from Ref.~\cite{Bus12} and it shares with them a number of essential physical features, the results obtained in Ref.~\cite{Kva21} also speak in favour of the ability of our model to handle cognitive dissonance and, potentially, other psychological phenomena. Interestingly, in Ref.~\cite{Kva21} only a combination of the classical and quantum models could satisfactorily reproduce alternations between the mental states. Since our model also combines the elements of a classical and quantum oscillators, a hybrid nature of our model and of that from Ref.~\cite{Kva21} confirms that advanced physics-focused quantum-mechanical approaches are needed to adequately describe complex psychological phenomena that involve oscillations of mental states.

\subsection*{Connection to Quantum Zeno models}
The quantum Zeno effect (QZE) has been employed to model bistable perception \cite{Atm04, Atm10, Atm13, Kor17}. This section explores the relationship between the proposed quantum tunnelling model and broadly understood Zeno models, examining how these approaches may complement each other in describing perceptual and cognitive processes. An earlier attempt to compare a quantum oscillator model of bistable perception with the QZE model was made in Ref.~\cite{Bus12}. The interested reader may also refer to the relevant work Ref.~\cite{yearsley2016zeno}.

The QZE \cite{Mis77, Ahm22} is a quantum mechanical phenomenon, where frequent observations of a quantum system inhibit the system's evolution, effectively ``freezing'' its state. Often referred to as a paradox, it suggests that a measurement, usually aimed to obtain information about a system, can instead prevent the system from changing states. This behaviour is sometimes described by the metaphor ``a watched pot never boils'' \cite{Atm04} that illustrates the effect in the limit of continuous observation.

The QZE has been widely investigated in the context of quantum optics \cite{Nak01, Rai12, Vys22} and quantum information processing \cite{Harr17, Paz12, Kum23, Abb24}. One of the important applications of the QZE is its potential to suppress the effect of quantum tunnelling \cite{Por22}, which has been demonstrated in an experiment involving the tunnelling behaviour of specially prepared atoms \cite{Pat15}. By cooling a gas of such atoms in a vacuum chamber and suspending them between laser beams, it was possible to control and image the tunnelling process. The findings revealed that the atoms tunnelled freely when the imaging laser was either off or the intensity of the light emitted by it was low. However, as the intensity of the laser beam and the frequency of measurements increased, the tunnelling events decreased significantly by virtue of the QZE.

The fundamental principles and original findings discussed earlier in this paper suggest a compelling connection between quantum mechanics and human behaviour. Just as quantum systems reveal only certain properties when measured, an individual's behaviour can vary significantly across different situations, revealing only specific characteristics in particular contexts \cite{Pat11, Mak24_information, Mak24_information1}. This perspective aligns with the idea that human behaviour comprises a spectrum of potential mental states, each state contributing to an individual’s overall behavioural profile \cite{Khr06, Bus12, Pot22}. However, in any given situation, only certain aspects of this behaviour become observable. Designing an experiment to highlight one behavioural trait may, consequently, reduce the accuracy with which other traits are captured or even render them invisible \cite{Fle10}. This principle also extends to our understanding of perception, where observing one aspect can obscure or alter the visibility of others, much like measurement in quantum systems.

The impact of the environment on behaviour has long been discussed in philosophy, notably in Ortega y Gasset’s assertion, ``I am me and my circumstance'' \cite{Ortega_Gasset}. More recent studies have drawn parallels between Ortega y Gasset’s notion of circumstance and quantum aspects of cognition \cite{deC13, Mak24_information}. Representing behaviour as a quantum system also aligns with the concept of discrete mental states \cite{Aer22, Aer22_1} and the interpretation of information as energy states within a physical system \cite{Dit14}.
\begin{figure}
 \includegraphics[width=0.99\textwidth]{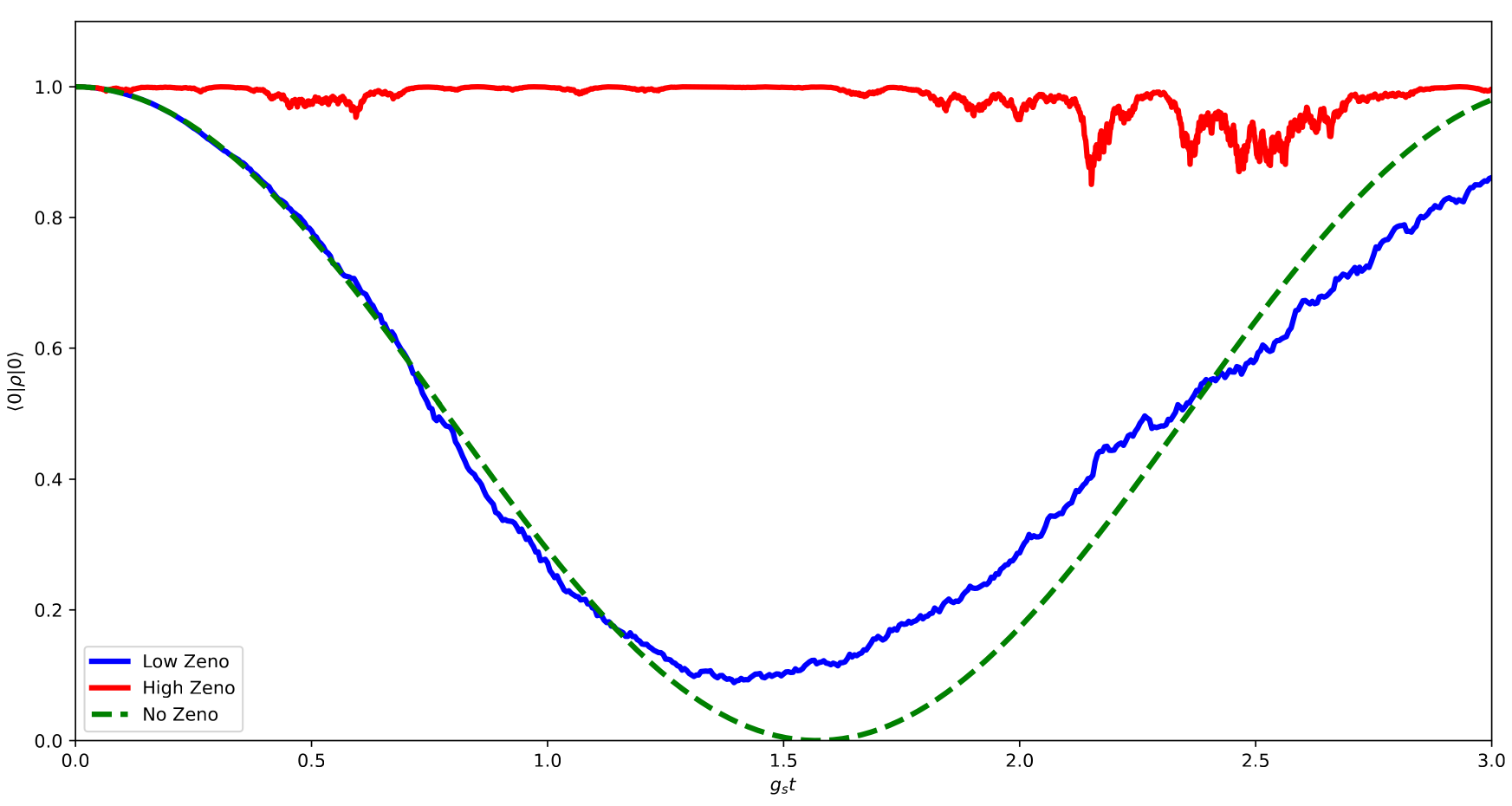}
 \caption{Illustration of the impact of the QZE on the oscillation dynamics of an idealised two-level quantum system.\label{QZE}}
\end{figure}

Thus, it is plausible to assume that the frequency of observing human behaviour has a similar effect on the outcome as changing the frequency of observation in a quantum system. Based on common sense, in the realm of human behaviour, frequent observations and feedback might similarly influence actions, potentially stabilising them in specific ways. For instance, if someone knows they are being monitored frequently, they may adjust or stabilise their actions to meet expected norms and avoid negative consequences \cite{Cla14}. This is analogous to the effect of continuous measurement on the evolution of a quantum system, where frequent ``observations'' constrain the dynamics of the system.

The same conclusion can be applied to the perception of ambiguous figures. As discussed throughout this paper and in the references cited above, deliberate interference with the natural frequency of eye blinks, as well as attempts to manipulate the perception of optical illusions, leads to changes in how these figures are perceived. This provides a link between optical illusions and QZE-based theories of quantum cognition and perception \cite{Atm04, Atm10}.

Since mathematical models of the QZE-controlled tunnelling effect are complex and involved, we chose an idealised yet conceptually similar two-level quantum oscillatory system to demonstrate the existence of such a link. This system can be described using a relatively straightforward theoretical formalism \cite{Jac06, Nie08, Abb24}. The connection between this formalism and the Schr{\"o}dinger equation-based model adopted in this paper is detailed in Ref.~\cite{Cam24}. This model is conveniently implemented as part of the standard {\tt QuantumOptics} module in the programming language Julia \cite{Julia}. For the sake of self-consistency, the key equations of the model are reproduced in the Methods section of this paper.

As shown in Fig.\ref{QZE}, when the system is not subject to measurement, it undergoes harmonic oscillation as a function of time. However, once measurement noise is introduced (as indicated by the curve labelled ‘Low Zeno’ in Fig.\ref{QZE}), the oscillatory behaviour is partially inhibited. Importantly, as the frequency of measurement increases (‘High Zeno’), the oscillation is suppressed.

This idealised model demonstrates how the QZE can be integrated into the more complex quantum tunnelling model proposed in this paper, with predictions that can be interpreted as follows. When visual attention, or ``measurement'', is frequently directed toward a particular interpretation of an ambiguous figure, such as the Necker cube, perception is less likely to switch to an alternative interpretation. Conversely, as attention relaxes, perception becomes more likely to alternate between the interpretations. Combined with the quantum tunnelling model, QZE-based theoretical frameworks may be extended to broader cognitive processes, suggesting that the mathematical apparatus of quantum mechanics can offer valuable insights into how repeated cognitive focus influences perceptual stability and transitions between mental states.

Finally, it is worth mentioning the existence of additional significant quantum-mechanical processes that can be employed to model mental systems and perceptual states accurately. These include violations of Bell-type inequalities, interference effects and certain alternatives to the conventional rules of probability theory \cite{Atm10, Atm13, Wad23}. Inequalities of information play a fundamental role in information theory and have been exploited effectively to establish bounds on the optimal rates of various information-processing tasks \cite{Cao16}. However, although considering these and potentially other quantum-mechanical effects in the context of the proposed model appears to be plausible  \cite{Kli21}, undertaking this endeavour lies well beyond the scope of the present work.

{\Ivan The discussion in the previous paragraph also highlights an important potential connection between the concept of quantum contextuality \cite{Koc67, How14} and the models presented in this paper. While quantum contextuality plays a significant role in quantum cognition theory \cite{Asa14}, the model proposed in this work, as well as the related works \cite{Mak24_APL}, does not rely on this concept. This is because, although both phenomena arise from the principles of quantum mechanics, they address distinct aspects. Quantum contextuality focuses on the dependence of measurement outcomes on the experimental setup, emphasising the non-classical nature of quantum properties. In turn, quantum tunnelling involves the probabilistic traversal of energy barriers, illustrating wave-particle duality and the non-deterministic behaviour of quantum particles. However, understanding both phenomena is essential for advancing quantum technologies and decision-making theory, and an attempt to explore a connection between them in the context of the studies presented in this paper has been done elsewhere~\cite{Mak24_book}. }

\subsection*{Practical Applications}
The proposed quantum-mechanical model of bistable perception offers a groundbreaking framework for practical applications across multiple fields. By connecting quantum principles to cognitive science, this model not only advances theoretical understanding but also provides tangible solutions for real-world challenges. Specifically, it has the potential to significantly impact domains such as video games, cognitive therapy, artificial intelligence applications, education and neuromarketing.

In the domain of video games, the model could revolutionise how perceptual ambiguities are explored and employed in interactive environments. Video games, with their inherent capacity for immersion and controlled experimentation, provide an ideal medium for testing the dynamics of bistable perception. For instance, games such as Superliminal \cite{game1} and Neckerworld \cite{Neckerworld} already integrate perceptual ambiguities, offering a fertile ground for integrating quantum principles (e.g.,~Ref.~\cite{wang2021game}). By dynamically adapting game play based on player responses---monitored through eye-tracking or brain activity---the model can enhance game play experiences while simultaneously generating valuable cognitive data. This approach could also lead to entirely new genres of games, where the mechanics themselves are inspired by quantum superposition, engaging players in unique cognitive challenges while advancing our understanding of perceptual dynamics.

The applications of the model extend beyond entertainment into the realm of cognitive therapy and rehabilitation. Disorders such as schizophrenia, autism and post-stroke impairments often involve challenges with perception and decision-making under uncertainty \cite{Zha15}. Tools inspired by this model, such as virtual reality (VR) environments or therapeutic games, could be designed to train individuals in stabilising their perceptual responses. For example, patients could interact with optical illusions or cognitive puzzles tailored to improve neural flexibility. These tools would monitor real-time perceptual states using biofeedback, dynamically adjusting interventions to the patient's progress. Such innovations align with existing research showing that cognitive training through interactive media can significantly improve neuroplasticity and behavioural outcomes \cite{bavelier2012brain}.

Artificial intelligence systems stand to benefit greatly from incorporating quantum-inspired decision-making models (e.g.,~Ref.~\cite{Pot22}). Current AI algorithms often struggle to replicate human-like reasoning in ambiguous situations. By embedding the probabilistic framework of quantum mechanics, our model can enhance AI's capacity to predict and adapt to human behaviour. For example, virtual assistants and educational platforms could use the model to interpret user intentions more effectively, particularly in scenarios involving conflicting inputs. In autonomous vehicles, the model could improve safety by predicting driver decisions during moments of uncertainty or distraction \cite{Abb24_1}. These applications illustrate how quantum-inspired cognition can help AI systems better mirror and interact with human users.

In education, the model provides an opportunity to develop innovative tools for teaching complex concepts. Gamified learning platforms could simulate quantum principles through puzzles or interactive lessons, introducing students to topics like superposition and cognitive biases (e.g.,~Ref.~\cite{cooper2010foldit, Neckerworld}). Additionally, professional training programs for high-stakes fields, such as aviation or surgery, could use VR simulations to enhance decision-making under pressure \cite{Cle13, Cle17}. By modelling bistable perception dynamics, these programs could help trainees develop greater cognitive flexibility and situational awareness, ultimately improving performance in critical tasks.

Neuromarketing is another area where the potential of the model is profound (e.g.,~Ref.~\cite{Khr14}). Perceptual ambiguity is often leveraged in marketing to capture consumer attention, as seen in Rubin's vase or similar visual stimuli \cite{Mak24_APL}. Using our model, practitioners could analyse how consumers resolve such ambiguities, tracking perceptual reversals through eye-tracking and EEG data. These insights could inform the design of advertisements and retail experiences that guide decision-making more effectively. The ability to predict consumer behaviour in ambiguous contexts could transform how brands engage with audiences, leading to more personalised and effective marketing strategies.

Beyond these specific applications, the broader implications of the model span disciplines. In social psychology, the model could be used to explore cognitive dissonance, where conflicting beliefs create mental ``barriers'' analogous to quantum tunnelling \cite{festinger1962dissonance}. Neuroscience could leverage the framework to investigate how neural circuits process ambiguous stimuli, shedding light on the biological basis of perception and decision-making \cite{Pia17}. Behavioural economics could also benefit by using the model to analyse anomalies in human choices, such as the Ellsberg paradox, providing a quantum-inspired perspective on risk and uncertainty \cite{blavatskyy2010models}.

\section*{Conclusion}
Our study presents a novel quantum-physical model that successfully integrates the principles of quantum mechanics with the physiological dynamics of bistable perception, specifically in relation to eye blink dynamics. The model proposes a mechanism for perceptual reversals, featuring energy conversion and redistribution processes, underscored by wave function interference within a quantum-mechanical context. By combining quantum mechanics with the dynamics of bistable perception, we also highlight how video games serve as a novel testing ground for exploring cognitive paradoxes and decision-making anomalies, paving the way for applications in both research and game development.

The proposed model assigns a key role to phase variations, particularly in relation to wave interference. Intriguingly, these phase shifts are linked with eye blinks, a function that refreshes eye lubrication and that has been found to impact perceptual state changes and object recognition. While existing research works have not established yet a definitive link between the optical properties of the lubricated eye and bistable perception, our findings suggest a plausible association, strengthened by the evidence of age-related declines in eye lubrication and blink rate coinciding with a decreased ability to experience bistable stimuli \cite{Sha14, lo2011investigation}. Our insights, capable of explaining perceptional reversals in video games, prompt a natural extension towards the integration of quantum principles into video game design and use of video games as a therapeutic treatment \cite{Li15}. Visual illusions, when embedded within a gameplay, could potentially serve as a powerful tool for probing the depths of human decision-making, offering a unique blend of entertainment and cognitive experimentation---a direction that is currently actively explored by computer scientists \cite{wang2021game}.

If we consider the conception of video games that intentionally incorporate visual illusions, strategically guiding players through scenarios where quantum-like behaviours emerge, these games could present players with choices that defy classical logic, mirroring the superposition of quantum states. By immersing players in environments where their decisions lead to outcomes that deviate from classical expectations, these games could serve as virtual laboratories for studying the nuances of decision-making under novel paradigms.

Such video games could offer controlled yet immersive environments for investigating the influence of quantum-inspired models on human behaviour. By designing game mechanics that challenge players' intuitive understanding of classical logic, developers could systematically collect data on decision patterns and responses. The interplay between visual illusions, quantum principles and player choices could illuminate new facets of decision-making dynamics, potentially revealing insights that transcend traditional experimental setups. Moreover, the application of elements of game playing in the domains of quantum physics and visual illusions has the potential to engage a broader audience, thereby democratising participation in research endeavours. For example, players, driven by the intrigue of deciphering quantum-like puzzles and navigating perceptual ambiguities, could contribute to large-scale data collection efforts. This collaborative approach not only enriches the research landscape but also promotes public understanding of quantum concepts and psychological phenomena. On the other hand, the integration of quantum models and visual illusions into the domain of video games heralds a captivating synergy between entertainment and cognitive exploration. By designing games that leverage the power of superposition states and challenge classical decision-making norms, researchers could unlock novel insights into human behaviour. These virtual laboratories could reshape the future of psychological research, allowing us to peer into the enigmatic world of quantum cognition through the immersive lens of interactive entertainment.

Finally, the scope of our model extends beyond bistable perception of the Necker cube and similar ambiguous figures. In particular, we have provided considerable arguments in favour of the ability of our model to capture the effects of cognitive dissonance. This versatility, in conjunction with the bridging of quantum principles and physiological mechanisms, underscores the potential influence of our findings on the field.

\section*{Methods}
\subsection*{Numerical solution of the Schr{\"o}dinger equation}
We numerically solve the Schr{\"o}dinger equation Eq.~(\ref{eq:SE}) using a finite-difference time-domain (FDTD) method \cite{Sullivan}. We split the wave function $\psi(x, t)$ into the real and imaginary parts:
\begin{equation}
  \label{eq:Eq2}
 \psi(x, t)=\psi_{re}(x, t)+i\psi_{im}(x, t)\, 
\end{equation}
and rewrite Eq.~(\ref{eq:SE}) as
\begin{gather}
  \label{eq:Eq3}
  \frac{\partial \psi_{re}(x, t)}{\partial t} = -\frac{\hbar}{2m}\nabla^2 \psi_{im}(x, t)+\frac{1}{\hbar}V(x)\psi_{im}(x, t) \\ \nonumber
\frac{\partial \psi_{im}(x, t)}{\partial t} = \frac{\hbar}{2m}\nabla^2 \psi_{re}(x, t)-\frac{1}{\hbar}V(x)\psi_{re}(x, t)\,, 
\end{gather}
where $m\approx9.1093837\times10^{-31}$\,kg and $\hbar\approx1.054571817\times10^{-34}$\,J$\cdot$s. The coordinate $x$, time $t$ and wave function $\psi(x,t)$ are represented as discrete quantities using a spatially uniform mesh with the size $\Delta x$ and a temporal mesh with the size $\Delta t$. The $x$-coordinate becomes a vector of discrete elements $x_k = k \Delta x$, where $k = 1\dots N_x$ and $N_x$ is the number of nodes of the spatial mesh. Similarly, the discrete time instances are $t_n = n \Delta t$ with $n = 1\dots N_t$. The value of $\Delta t$ must be related to $\Delta x$ via the Courant stability criterion \cite{Sullivan}:
\begin{equation}
  \label{eq:Eq4}
  \Delta t = \frac{1}{8}\frac{2m}{\hbar} (\Delta x)^2\,. 
\end{equation}
Thus, we obtain a spatio-temporally discretised representation of Eq.~(\ref{eq:Eq3}): 
\begin{gather}
  \label{eq:Eq5}
  \psi_{re}^{n}(k) = \psi_{re}^{n-1}(k)-\frac{1}{8}\left[\psi_{im}^{n-1/2}(k+1)-2\psi_{im}^{n-1/2}(k)+\psi_{im}^{n-1/2}(k-1)\right]+\frac{\Delta t}{\hbar}V(k)\psi_{im}^{n-1/2}(k)\\ \nonumber
  \psi_{im}^{n}(k) = \psi_{im}^{n-1}(k)+\frac{1}{8}\left[\psi_{re}^{n-1/2}(k+1)-2\psi_{re}^{n-1/2}(k)+\psi_{re}^{n-1/2}(k-1)\right]-\frac{\Delta t}{\hbar}V(k)\psi_{re}^{n-1/2}(k)\,. 
\end{gather}

The electron is modelled as a Gaussian-shape energy wave packet formed before the start of the simulation at the discrete instant of time $n=0$:
\begin{gather}
  \label{eq:Eq6}
  \psi_{re}^{0}(k) = \exp\left(-0.5\left(\frac{k-k_0}{\sigma}\right)^2\right) \cos\left(\frac{2\pi(k-k_0)}{\lambda}\right)\\ \nonumber
  \psi_{im}^{0}(k) = \exp\left(-0.5\left(\frac{k-k_0}{\sigma}\right)^2\right) \sin\left(\frac{2\pi(k-k_0)}{\lambda}\right)\,, 
\end{gather}
where $\lambda$ is the wavelength, $\sigma$ is the width of the Gaussian pulse and $k_0$ is the coordinate of origin of the pulse. Since we require the electron to be present somewhere in the potential well, the amplitudes of the wave functions are normalised as
\begin{equation}
  \label{eq:Eq7}
  \int_{-\infty}^{\infty} \psi^{*}(x)\psi(x) \,dx = 1\,. 
\end{equation}
The probabilities of funding the electron is the $|0\rangle$ and $|1\rangle$ regions of the potential well are calculated as 
\begin{gather}
  \label{eq:Eq8}
  P_{|0\rangle}=\int_{-\infty}^{x_{centre}} \psi^{*}(x)\psi(x) \,dx \\ 
  P_{|1\rangle}=\int_{x_{centre}}^{\infty} \psi^{*}(x)\psi(x) \,dx\,, 
\end{gather}
where $P_{|0\rangle}+P_{|1\rangle}=1$. The following physically meaningful model parameters were used in the simulations: $\Delta x=0.1\times10^{-11}$\,m, $N_x=4001$, $N_t=3\times10^6$, $\lambda=1.6\times10^{-10}$\,m and $\sigma=1.6\times10^{-10}$\,m.

The chosen model parameters are often used in FDTD simulations of Gaussian-shape energy wave packets \cite{Sullivan}. They are also correlated with the experimental parameters used in studies of tunnelling of electrons in semiconductor double-barrier structures \cite{Cha74}. It is noteworthy that these parameters can be further adjusted to model data obtained in a perception reversal experiment. In this case, an optimal set of parameters will not necessarily corresponds to those of a real-life physical system, but it will be dictated by the fundamental limits of the Schr{\"o}dinger equation \cite{Gri04} and numerical limitations of the FDTD method \cite{Sullivan}.

\subsection*{Model of the Quantum Zeno Effect}
The QZE model adopted in this paper closely follows the general model presented in Ref.~\cite{Abb24}. A similar model is implemented in {\tt QuantumOptics} module of the programming language Julia~\cite{Julia}, which in turn builds on the theoretical formalism developed in Refs.~\cite{Jac06, Nie08}. 
 
Consider a qubit realised as a physical system consisting of a two-level atom trapped inside a cavity. Representing the qubit as a mode $\sigma$ and denoting the cavity mode as $a$, we can formulate a model that is entirely isolated from external influences, allowing only for measurement-controlled interactions between the qubit and a coherent field input, while also accounting for the inherent decay of the cavity.

Let the atom has two possible spin states $|0\rangle=|\uparrow\rangle$ and $|1\rangle=|\downarrow\rangle$ that interact with a quantised electromagnetic field within a cavity. The governing Hamiltonian $\hat{H}_i$ for this atom-cavity interaction can be written as 
\begin{eqnarray}
\hat{H}_i = g a^\dagger a \sigma_{-} \sigma_{+}\,,
\label{eq:RC2}
\end{eqnarray}
where $g$ is the strength of the atom-cavity coupling, $a$ is the cavity annihilation operator, and $\sigma_{-}$ and $\sigma_{+}$ are the lowering and raising operators for the atom, respectively. The interaction between the atom and the cavity results in an exchange of energy, giving rise to Rabi oscillations---oscillations of energy between the atom and the cavity.

A coherent driving of the system with an external signal that is incident on one of the cavity mirrors can be modelled by the Hamiltonian
\begin{eqnarray}
\hat{H}_c = -i\beta(a^\dagger - a)\,,
\label{eq:RC3}
\end{eqnarray}
where $\beta$ is the amplitude of the driving signal. The state of the atom undergoes continuous monitoring via a coherent measurement process, enabling real-time observations and subsequent control of the quantum state. These processes are incorporated into the Hamiltonian as
\begin{eqnarray}
\hat{H}_z = g_z (\sigma_{+} + \sigma_{-})\,,
\label{eq:RC4}
\end{eqnarray}
where $g_z$ is the amplitude of the coherent atomic driving. Since the atom collapses into an eigenstate depending on the measurement frequency, the choice of $g_z$ controls the dynamics of the system since it determines the rate at which the state of the atom is measured.

We analyse the dynamics of the system using a stochastic master equation approach \cite{Nie08}. This equation is analogous to the time-dependent Schr{\"o}dinger equation, incorporating both the atom-light interaction and the effect of frequent measurements on the atom's spin states. The full Hamiltonian for the system is given by
\begin{eqnarray}
\hat{H} = \hat{H}_i +\hat{H}_c +\hat{H}_z\,, 
\label{eq:RC6}
\end{eqnarray}
where $\hat{H}_i$, $\hat{H}_c$ and $\hat{H}_z$ are given by Eqs.~\eqref{eq:RC2},~\eqref{eq:RC3}~and~\eqref{eq:RC4}, respectively.
The time evolution of the density matrix $\rho$, the linear stochastic master equation of which accounts for the effects of decoherence and dissipation, can be written as
\begin{eqnarray}
\dot{\rho }=-i[\hat{H},\rho ]+\hat{C}\rho {\hat{C}}^{{\dagger} }-\frac{1}{2}{\hat{C}}^{{\dagger} }\hat{C}\rho -\frac{1}{2}\rho {\hat{C}}^{{\dagger} }\hat{C}\,,
\label{eq:RC7}
\end{eqnarray}
where $\hat{H}$ is given by Eq.~\eqref{eq:RC6} and $\hat{C} = \sqrt{\kappa} a$ is the collapse operator associated with the cavity decay and $\kappa$ is the decay rate. Adding a white noise term as shown in Ref.~\cite{Julia}, we numerically solve Eq.~\eqref{eq:RC7} following the algorithm presented in Ref.~\cite{Julia}, which enables us to simulate measurement-driven dynamics of the atom-cavity system.

\subsection*{Transmission Probability}
{\Ivan In this subsection, we highlight the nonlinear properties of the model proposed in this paper. The quantum tunnelling probability function is nonlinear in general. The probability of a particle tunnelling through a potential barrier is governed by the transmission coefficient $T$ that can be derived from the Schr{\"o}dinger equation. For a rectangular potential barrier, the transmission coefficient is given by
\begin{equation}
T = \dfrac{1}{1 + \dfrac{V_0^2}{4E(V_0 - E)} \sinh^2 (\gamma a)}\,,
\end{equation}
where $V_0$ is the height of the barrier, $E$ is the energy of the particle, $a$ is the width of the barrier and $\gamma = \sqrt{\dfrac{2m(V_0 - E)}{\hbar^2}}$ is the decay constant \cite{McQ97}.

Several factors contribute to the nonlinearity of the tunnelling probability, including exponential dependence as the transmission probability is a function of $\sinh^2$. Neuromorphic quantum neural networks design to model the perception of optical illusions by human agents exploit the effect of quantum tunnelling as a nonlinear activation function \cite{Mak24_APL}.}%Ivan

\bibliography{sample}

\section*{Funding}
The authors did not receive any funding to support this project.

\section*{Acknowledgements}
The authors would like to thank Dr.~A.~H.~Abbas for fruitful discussions that helped improve this work.

\section*{Data Availability}
The datasets generated and analysed in Study~1 are available in the GitHub repository, \url{https://github.com/BehaviouralDataScience/Quantum-Bistable-Perception}. Data produced in Study~2 and Study~3 are included in this article. 

\section*{Author contributions statement}
I.S.M and G.P.~conceived the project. G.P.~collected data and analysed results for Study~1. I.S.M.~conducted the calculations and analysed results for Studies 2 and 3. Both authors wrote and reviewed the manuscript. 

\section*{Competing interests}
The authors declare no competing interests.

\section*{Ethical approval and informed consent}
In this study, we recruited 78 participants to engage in the experimental tasks to explore decision-making and perceptual processes under controlled conditions. This study was approved by the Humanities and Social Sciences Research Ethics Committee (HSSREC) of the University of Warwick (approval number 31/13-14). All experiments were performed in accordance with relevant guidelines and regulations, including the ethical principles outlined in the Declaration of Helsinki. Upon arrival at the laboratory, each participant received a study information sheet and a consent form. Informed consent was obtained from all participants prior to their participation. All participants were adults (18 years or older), and no minors were involved in the study; therefore, parental or guardian consent was not required.

\end{document}